# Adaptive Tuning Algorithm for Performance tuning of Database Management System

S.F.Rodd[1], Dr. U.P.Kulkrani[2]

[1] Asst. Prof., Gogte Institute of Technology, Belgaum, Karnataka, INDIA
Email: sfrodd@rediffmail.com

[2] Prof., SDMCET Dharwar, Karnataka, INDIA.
Email: upkulkarni@yahoo.com

**Abstract** - Performance tuning of Database Management Systems(DBMS) is both complex and challenging as it involves identifying and altering several key performance tuning parameters. The quality of tuning and the extent of performance enhancement achieved greatly depends on the skill and experience of the Database Administrator(DBA). As neural networks have the ability to adapt to dynamically changing inputs and also their ability to learn makes them ideal candidates for employing them for tuning purpose. In this paper, a novel tunig algorithm based on neural network estimated tuning parameters is presented. The key performance indicators are proactively monitored and fed as input to the Neural Network and the trained network estimates the suitable size of the buffer cache, shared pool and redo log buffer size. The tuner alters these tuning parameters using the estimated values using a rate change computing algorithm. The preliminary results show that the proposed method is effective in improving the query response time for a variety of workload types.

## I. INTRODUCTION

Database Management Systems are an integral part of any corporate house, the online systems, and e-commerce applications. For these systems, to provide reliable services with quick query response times to their customers, the Database Management Systems(DBMS) must be functioning efficiently and should have built-in support for quick system recovery time in case of partial failure or system resource bottlenecks. The performance of these systems is affected by several factors. The important among them include database size which grows with its usage over a period of time, increased user base, sudden increase in the user processes, improperly or un-tuned DBMS. All of these tend to degrade the system response time and hence call for a system that anticipates performance degradation by carefully monitoring the system performance indicators and auto tune the system.

Maintaining a database of an enterprise involves considerable effort on part of a Database Administrator (DBA) as, it is a continuous process and requires in-depth knowledge, experience and expertise. The DBA has to monitor several system parameters and fine tune them to keep the system functioning smoothly in the event of reduced performance or partial failure. It is therefore desirable to build a system that can tune itself and relieve the DBA of the tedious and error prone task of tuning. Oracle 9i and 10g have built in support for tuning in the form of tuning advisor. The tuning advisor estimates the optimal values of the tuning parameters and recommends them to the DBA. A similar advisor is also available in SQL Server 2005 which is based on what-if analysis. In this approach, the DBA provides a physical design as input and the Tuning Advisor performs the analysis without actually materializing the physical design. However, the advisor available in 2005 recommends the changes needed at the physical level such as creation of index on tables or views, restructuring of tables, creation of clustered index etc. which are considered to be very expensive in terms of Database Server down time and the effort on part of the DBA.

## II. RELATED WORK

Several methods have been proposed that proactively monitor the system performance indicators analyze the symptoms and auto tune the DBMS to deliver enhanced performance. Use of Materialized views and Indexes, Pruning table and column sets[1-2], Use of self healing techniques[3-4], use of physical design tuning are among the proposed solutions. The classical control is modified and a three stage control involving Monitor, Analyze and Tune[6] is employed to ensure system stability. The architecture presented in [5] for self healing database forms the basis for the new architecture presented here in this paper. This paper presents a new DBMS architecture based on modular approach, where in each functional module can be monitored by set of monitoring hooks. These monitoring hooks are responsible for saving the current status information or a snapshot of the server to the log. This architecture has high monitoring overhead, due to the fact that when large number of parameters to be monitored, almost every module's status information has to be stored on to the log and if done frequently may eat up a lot of CPU time. Moreover, this architecture focuses more on healing the system and does not consider tuning the DBMS for performance improvement.

Ranking of various tuning parameters based on statistical analysis is presented in[6]. The ranking of parameters is based on the amount of impact they produce on the system performance for a given workload. A formal knowledge framework for self tuning database system is presented in[7] that defines several knowledge components. The knowledge components include Policy knowledge, Workload knowledge, Problem diagnosis knowledge, Problem Resolution Knowledge, Effector knowledge, and Dependency knowledge. The architecture presented in this paper involves extracting useful information from the system log and also from the DBMS using system related queries. This information gathered over a period of time is then used to train a Neural Network for a desired output response time. The neural network would then estimate the extent of correction to be applied to the key system parameters that help scale up the system performance.

### III. PERFORMANCE TUNING

Calibrating the system for desired response time is called performance tuning. The objective of this system is to analyze the DBMS system log file and apply information extraction techniques and also gather key system parameters like buffer miss ratio, number of active processes and the tables that are showing signs of rapid growth. The control architecture presented in this paper, only one parameter namely, the buffer cache is tuned. Using the statistical information of these three parameters to train the Neural Network and generate an output that gives an estimate of the optimal system buffer size. Since, the DBMS are dynamic and continuously running around the clock, the above information must be extracted without causing any significant system overhead.

Extracting information from system log ensures that there is no overhead. The queries that are used to estimate buffer miss ratio, table size and number of user processes are carefully timed and their frequency is adjusted so that it does not add to the overhead in monitoring the system.

### IV NEURAL NETWORK

As neural networks are best suited to handle complex systems and also have ability to learn based on the trained data set, the same is used in the proposed architecture. As shown in Fig. 1, Neural Network will have p inputs, a specified number of nodes in the hidden layer and one or more output nodes. The neural network used in this control architecture is a feed forward network. The activation function used is sigmoid function. It is this function that gives the neural network the ability to learn and produce an output for which it is not trained. However, the neural networks need a well defined training data set for their proper functioning.

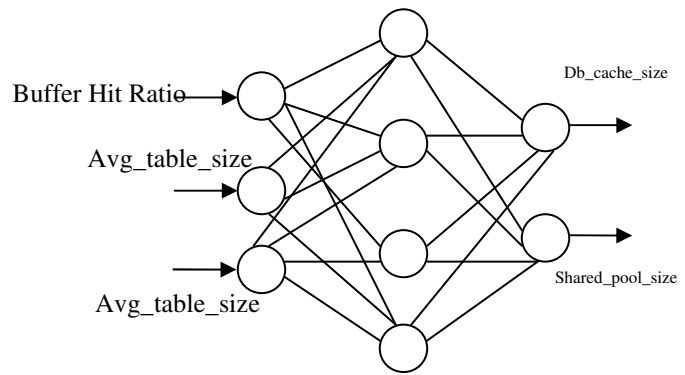

Figure 1.  Basic Neural Network Structure

The neural networks work in phases. In the first phase, the network is trained using a well defined training set for a desired output. In the second phase a new input is presented to the network that may or may not be part of the training data set and network produces a meaningful output. For the proper working of the neural network, it is important to choose a proper activation function, learning rate, number of training loops and sizeable number of nodes in the hidden layer.

### V. PROPOSED ARCHITECTURE

Fig. 2 Shows the architecture employed for identifying the symptoms and altering key system parameters. The DBMS system log file will be the primary source of information that helps in checking the health of the system. Since, the log file contains huge of amount of data, the data is first compressed into smaller information base by using a data mining tool. The architecture has Data Miner, Neural Network aggregator and Tuner as the basic building blocks. After, extracting meaningful information, the next step is to estimate the extent of correction required.

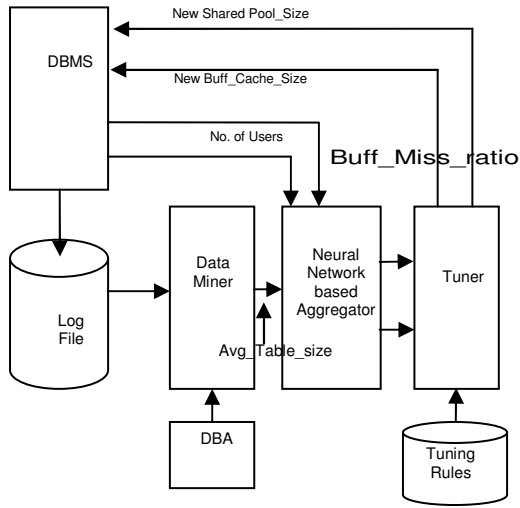

Figure 2. Neural Network based Tuning Architecture

As suggested in[2] physical tuning should be avoided as it is expensive. Most importantly the internal corrective measure such as altering the buffer size of the DBMS used in query processing is explored in this architecture. However, several parameters can be altered simultaneously for better performance gain. The Neural network estimates the required buffer size based on the current DBMS input parameters and the tuner applies the necessary correction to the buffer size based on the tuning rules. The tuner triggers a control action to fine tune the performance of the dbms based on the following algorithm

**ALGORITHM** dbTune(ESTMTD_DB_CACHE_SZ)
**Begin**

  Compute the change in response time since
  last modification  (ΔRtime)
 If ( ΔRtime >0 and ΔRtime > Rth)
      Increase the new buffer_size to next
      higher granule size
      Issue a command to alter the dbcache size
      to the new value
  Else
      If(ΔRtime <0 and ΔRtime < Rth)

      Decrease the new buffer size to next lower
      granule size.
      Issue a command to alter the dbcache size
      to the new value
**End**

## VI. EXPERIMENTAL RESULT

Table I shows the sample training data. A training data set of size 100 was used to train the Neural Network. As can be seen from the table, the buffer size is adjusted for increased table size, Number of user processes and Buffer Miss Ratio so that query execution time is reduced and the memory is used efficiently.

| Tab. Size (in no. of records) | Buff.Miss Ratio | Shared Pool size (in MB) | Buff. Size (in MB) |
|---|---|---|---|
| 1000 | 0.9824 | 32 | 4 |
| 1000 | 0.9895 | 32 | 4 |
| 1000 | 0.9894 | 32 | 8 |
| 1000 | 0.9505 | 32 | 8 |
| 2000 | 0.947 | 32 | 8 |
| 2000 | 0.9053 | 40 | 8 |
| 2000 | 0.8917 | 40 | 16 |
| 2500 | 0.875 | 40 | 16 |

Table I. Sample Training Data Set

The experiment was carried on Oracle 9i with a 3-input 2-output feed forward neural network with 100 internal nodes. The training is carried with an epoch value of 100, learning rate of 0.4 and with a training dataset of size 100. The estimated buffer size generated by the Neural Network is based on the dynamic values of the above three parameters as input. The tuner takes this input and alters the buffer size. The results obtained are really promising. As can be seen from the output in Fig. 4 the execution time is significantly lower for the increasing value of the buffer size. The query used takes join of three tables and generate huge dataset as result.

Fig. 3 shows the effect of buffer cache size on the query response time. TPC-C type benchmark load was used which represents an OLTP type load. As number of users grow beyond 12, the query response time starts rises rapidly. This is sensed by the neural network and calculates an appropriate size of the new buffer size. The tuner uses the tuning rules to apply the required correction. The tuning rules indicate when and at what interval of the buffer size, the correction is to be applied. Tuning the DBMS frequently may affect the performance and also lead to instability.

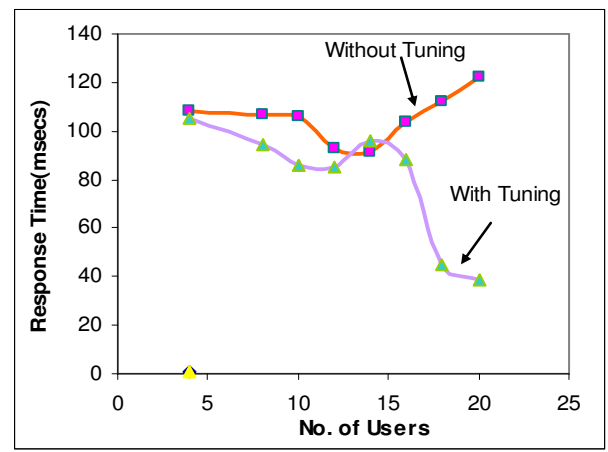

Figure 3. Effect of Buffer size on Query Response Time

## VII. CONCLUSION

A new tuning algorithm is presented in this paper. The Neural Network estimates the buffer cache size based on the trained data set. The correction is applied in accordance with the tuning algorithm so as to scale up system performance. This architecture learns from a training set to fine tune the system and thus it relieves the DBA of the tedious process of tuning the DBMS and also need for an expert DBA. Monitoring the macroscopic performance indicators ensures that there is little monitoring overhead. However, the system needs further refinement that takes into account sudden surge in work loads and also the neural network training dataset must be derived based on proper database characterization. It is also important to ensure that the system remains stable and gives consistent performance over a long period of time.

ACKNOWLEDGEMENTS

We would like to thank Prof. D.A.Kulkarni for scruitimizing the paper and for his valueable suggestions. Special thanks to Prof. Santosh Saraf for his help in learning Neural Network implementation in MATLAB. We extend our thanks to Computer Center, GIT, for providing laboratory facilities. We thank our esteemed Management for their financial support.